\documentstyle[11pt,newpasp,twoside,epsf]{article}

\def\etal   {{\it et al.}~}
\def\lya    {Ly$\alpha$~}

\def\nh     {N$_{\rm H}$~}
\def\chandra {{\it Chandra}~}

\def\edcomment#1{\iffalse\marginpar{\raggedright\sl#1\/}\else\relax\fi}
\reversemarginpar

\begin{document}

\title{Recent \bf{\it Chandra} Results on AGNs \& Future Prospects}
\author{Smita Mathur}
\affil{The Ohio State University, 140 West 18th Ave., Columbus, OH 43210}

\begin{abstract}
I review {\it Chandra}'s achievements in the last two years in AGN
research. I concentrate on some topics of my interest; some others are
reviewed in an accompanying article by K. Nandra. I comment briefly on
the need to allow for the discovery space to ensure rapid progress in
the field.
\end{abstract}

\section{\bf{\it Chandra} At Sharp Focus}

This meeting is called ``{\it Chandra} At Sharp Focus''. So let me
begin at the beginning when {\it Chandra} was being focused. The
radio-loud quasar PKS~0637-752, known to be a ``point source'', was
observed to achieve the sharp focus. What was detected is shown in
figure 1. The very first observation with \chandra led to the
discovery of a 100 kpc jet in this quasar, immediately showing the
unprecedented power of the observatory (Schwartz \etal 2000). Many
more observations of X-ray jets then followed (Marshall \etal 2001,
Sambruna \etal 2001). These are not just pretty
pictures. Multiwavelength observations, radio--optical--X-ray, allowed
determination of spectral energy distributions to constrain the
emission mechanisms.

Observations revealed that the jets are complex. In most cases, the
overall jet morphology in different wavebands was remarkably similar
(fig. 1), but there were subtle differences. In some cases (e.g. 3C
273), simple Synchrotron models provide good fit to the data, but in
some cases they do not. In PKS 0637-752, Synchrotron self-Compton
models imply extreme departure from equipartition of magnetic and
particle energy, while the models invoking the inverse-Compton of
microwave background photons require large Doppler factors ($\delta
\sim 10$, Tavecchio \etal 2000).

\begin{figure}
\caption{\chandra image of the X-ray jet in PKS 0637-752 (from Schwartz \etal 2000). Radio contours are overlayed on the X-ray image.}
\vspace{3.0truein}
\plotfiddle{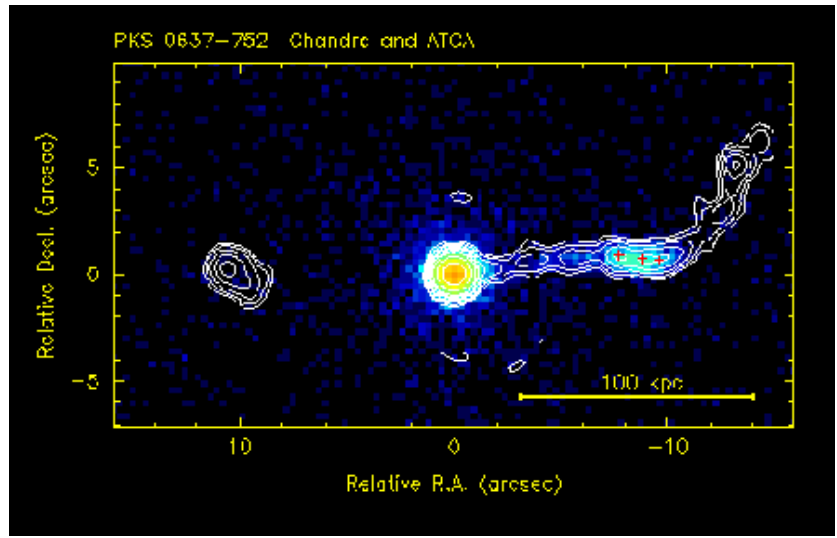}{2.5in}{0.0}{70}{70}{-200.0}{0.0}
\vspace{-2.5truein}
\end{figure}

Because of these complexities, the new \chandra observations are often
viewed as posing problems to our understanding of jet physics. I do not
view them as problems; they in fact offer opportunities to understand
particle acceleration and magnetic field structure. In this respect it
is instructive to see what Begelman, Blandford and Rees wrote in 1984:
``With AXAF (now renamed \chandra), it should be possible to detect
X-ray emission from several more jets.  ....it will then be possible
to make improved estimates of the magnetic field strengths, and so
\underline{test} the equipartition hypothesis''. I think the important
word here is ``test''.

From this demonstration of the excellent imaging capability of
\chandra, now I will move on to science exploiting its unprecedented
spectroscopic quality.

\section{Warm Absorbing Outflows in AGNs}

\subsection{Pre-\bf{\chandra} Era}

Absorption edges due to OVII/OVIII were detected in several AGNs with
ROSAT (e.g. Fiore \etal 1993) and ASCA (e.g. George \etal
1998). Associated UV absorption lines due to OVI, CIV, NV and \lya
were also detected in several AGNs (e.g. Bahcall \etal 1993). In
Seyfert galaxies, one-to-one correspondence was observed between the
UV and X-ray absorbers (Crenshaw \& Kraemer 1999). Photoionization
models then allowed determination of physical conditions in the
absorbing gas. Over the years (1994--2001), Mathur \& collaborators
have argued that the X/UV absorber must be highly ionized, high
column density, outflowing gas located at/outside the broad emission
line region of AGNs. Note that the information that the X-ray warm
absorbing gas is ``outflowing'' came only with its association with
the UV absorber. The important point is that over a range of
parameters, a photoionized plasma would imprint signatures in both
X-ray and UV bands. Obviously, if the gas is too highly ionized, there
may not be detectable UV absorption lines and if the gas has too low
column density then there would not be any detectable X-ray
absorption. Complexity, as commonly seen in UV absorption systems is
also expected in X-ray warm absorbers (see Mathur 1997 and references
there in). Understanding the physical conditions in warm absorbers is
important because the outflow seems to carry a significant amount of
kinetic energy and the mass outflow rate, in many cases, is comparable
to the accretion rate needed to power the AGN (Mathur \etal 1995).

\subsection{Warm Absorber models: Predictions}

The above understanding of the physical conditions in the warm
absorbers led to following predictions. 1) In addition to the edges,
resonance absorption lines due to highly ionized elements
should also be present (Nicastro \etal 1999). 2) The X-ray absorption
lines should show blue-shifts, similar to those in the UV
lines. 3) FWHM of X-ray lines should match the UV lines. 4) The
column densities of ions derived from the X-ray and UV lines should
match. Again, because of the exceptions discussed above, not {\it
every} absorption system in X-rays would have a match in the UV and
vice-a-versa. 

\subsection{\bf{\chandra} Observations of Warm Absorbers}

High resolution spectroscopy of AGNs known to have warm absorbers,
was performed in the first two years of \chandra. Below I review four
such cases. \\

\begin{figure}
\caption{LETG/HRC Spectrum of NGC~5548 (from Kaastra \etal 2000).}
\vspace{-2.6truein}
\plotfiddle{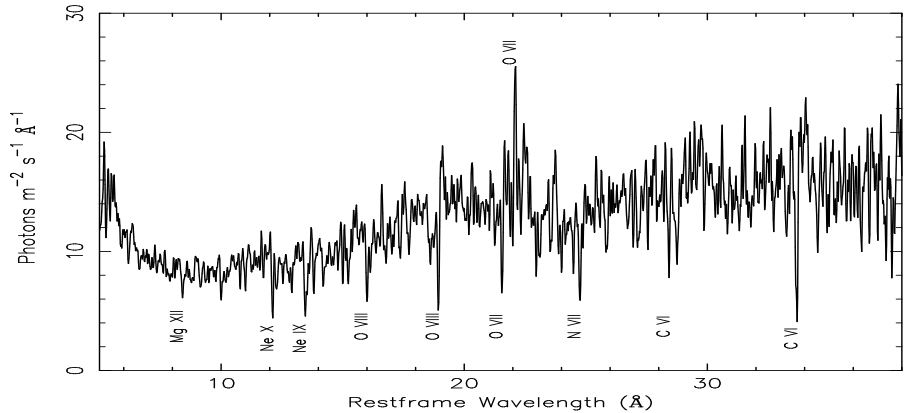}{2.5in}{-90.0}{50}{30}{-170.0}{0.0}
\vspace{2.6truein}
\end{figure}

\noindent
{\it NGC 5548}: This well studies AGN was observed with
LETG/HRC. Figure 2 shows the rich absorption line spectrum. Strong,
narrow absorption lines from highly ionized species are clearly
present (Kaastra \etal 2000). The widths of the X-ray lines are
consistent with the UV lines. The average blueshift of the X-ray
absorption lines was found to be somewhat smaller, but comparable to
the UV lines. However, further relation between the X-ray and UV
absorbers remained an open issue because of the low S/N and lower
resolution of the X-ray spectrum. (The spectral resolution of LETG is
$\sim$300--1000 compared to $\sim$10,000 of HST/STIS in medium
resolution.)\\

\noindent
{\it NGC 3783} was observed with HETG/ACIS-S (Kaspi \etal 2000,
2001). The high resolution spectrum revealed the presence of narrow
absorption lines of H- and he-like ions of O, Ne, Mg, Si, S and Ar as
well as FeXVII--FeXXI L-shell lines. These observations showed that
the warm absorber is indeed outflowing with blueshifts of the lines
consistent with the UV lines. The velocity dispersion was also found
to be consistent with the UV lines. In the first paper (2000) the
authors claimed that the column densities of X-ray and UV absorption
lines do not match. This was a surprise given the consistency of
Chandra and ASCA observations and the excellent ASCA/UV match shown by
Shields \& Hamann (1997). The predicted UV line strengths are highly
sensitive to the assumed continuum shape in the unobservable EUV
range. Indeed, in the second paper (2001), the authors found this
sensitivity to be case and the column densities predicted from the
low-ionization component were consistent with the UV lines, given a
particular EUV slope. \\

\noindent
{\it NGC 4051}: This narrow line Seyfert 1 galaxy was observed
simultaneously with \chandra (HETG/ACIS-S, spectral resolution
$\sim$1000) and HST (STIS) (Collinge \etal 2001). The authors have done
an excellent job of analyzing the data and have been careful in their
interpretation. Again, a number of absorption lines were
detected. Photoionization modeling revealed two separate components: a
high ionization system with high outflow velocity and a low ionization
system with low velocity. The UV observations also detected absorption
lines, as expected, with multiple velocity components. The
low-velocity X-ray absorption is consistent in velocity with many of
the UV systems, but the high velocity X-ray absorption does not seem
to have any UV component. The authors point out that X-ray
spectroscopy with even higher resolution (matching that of STIS) may
show multiple components in X-ray lines as well. \\

\begin{figure}
\caption{HETG/ACIS-S Spectrum of MCG-6-30-15 with a warm absorber
model overplotted (from Lee \etal 2001).}
\vspace{1.5truein}
\plotfiddle{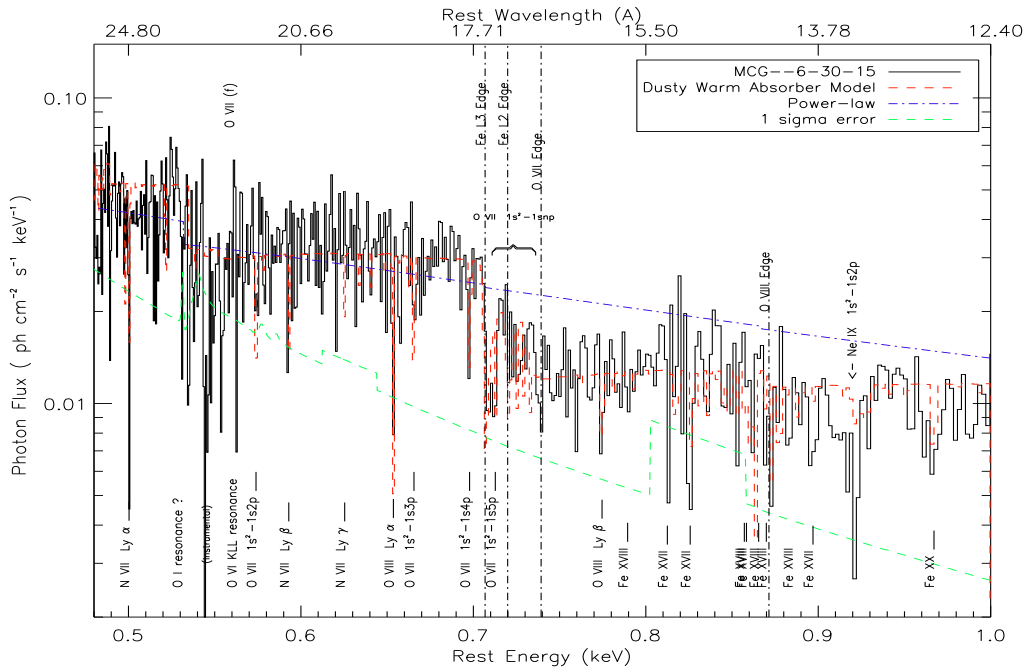}{2.5in}{90.0}{60}{50}{250.0}{0.0}
\vspace{-0.5truein}
\end{figure}

\noindent
{\it MCG-6-30-15}: This AGN, well-known from ASCA for its broad Fe
K$\alpha$ line, was known to harbor a warm absorber with possibly two
components (Reynolds \etal 1995). XMM-Newton RGS observations of this
source, however, lead Branduari-Raymont \etal (2001) to propose a
radically alternative explanation for the soft X-ray features. These
authors claimed that there is no warm absorber in MCG-6-30-15, and the
soft X-ray features are best explained by relativistically broadened
emission lines. The higher resolution \chandra HETG/ACIS-S
observation, therefore, were particularly interesting and demonstrated
the superior quality of \chandra gratings. The \chandra observations
detected a numerous absorption lines due to highly ionized species of
a number of elements (Figure 3). In a detailed analysis, Lee \etal
(2001) showed that a warm absorber model is not only adequate for
MCG-6-30-15, the data require it. These observations also revealed
first clear detection of OVI KLL resonance absorption, recently
predicted by Pradhan (2000). It is interesting too see how the
progress in high resolution X-ray spectroscopy is going hand in hand
with atomic physics. \\

\noindent
Thus, all the above observations of warm absorbers have shown
consistency with the model predictions (\S 2.2) based on a common
origin of the X-ray and UV absorbing gas.

\section{Broad Absorption Line Quasars (BALQSOs)}

The UV properties of BALQSOs are similar to the associated absorption
systems discussed above, except that their absorption lines are very
broad, with terminal outflow velocities up to 0.1c. Therefore it is very
important to understand the physical conditions in the absorbing gas as
the outflow might imply an energy budget problem. The physical
conditions are poorly constrained by the UV line studies alone and
X-ray spectra would potentially provide complementary information as
demonstrated by the X/UV absorbers discussed above. However, BALQSOs
are markedly underluminous in soft X-rays, with most observations
resulting in non-detections. Recent studies (Mathur \etal 2000, 2001,
Gallagher \etal 2001a and references there in) imply that the BALQSOs
are not intrinsically X-ray weak, but the strong absorption makes them
appear faint. In some cases there is clear evidence of partial
covering by the absorber and at least in one case there is indication
of a steep X-ray power-law slope. If true, this is extremely important
as it bears on the evolutionary scenarios of quasars (Mathur 2000).

\subsection{\bf{\chandra} observations of BALQSOs}

The sharp point spread function and very low background of \chandra
make it ideal for detecting faint point sources like BALQSOs. We
carried out a snap-shot survey of 10 BALQSOs with \chandra, and
detected 8 of them (Green \etal 2001). Each source had too few counts
to do meaningful spectroscopy, so a composite spectrum was generated
by stacking them all together. The best fit model was an absorbed
power-law (\nh = 2--10 $\times 10^{22}$ cm$^{-2}$) partially covering
the source. The best fit slope is similar to that of non-BAL
radio-quiet quasars, showing clearly for the first time that BALQSOs
are not intrinsically different from non-BAL QSOs. Note, however, that
the large uncertainty in fitted parameters allows for a steep
spectrum. Correcting for absorption, the X-ray luminosity of
high-ionization BALQSOs was found to be similar to that of normal
quasars. The low-ionization BALQSOs, however, were still underluminous
in X-rays. So the low-ionization BALQSOs are clearly different from
the high-ionization BALQSOs: they either have higher column density
absorbers, have steeper spectra, or are intrinsically fainter. Either
of the latter two possibilities would make them intrinsically
different from non-BAL QSOs. Preliminary analysis from a separate
\chandra program on BALQSOs is reported in Gallagher \etal (2001b).

\section{Emission Lines from AGNs}

The absorption lines discussed in $\S 2$ are observed when there is
absorbing gas along the line of sight. The same plasma, when viewed
from another angle should exhibit emission lines. In addition to the
resonance lines, seen in absorption, forbidden and intercombination
lines should also be present in emission (see also Krolik \& Kriss
1995). Such emission lines are best studied when the bright nuclear
continuum is suppressed. Naturally, the first observations to study
emission lines from AGNs were of highly absorbed Seyfert
galaxies. Below I review two such examples. \\

\begin{figure}
\caption{HETG/ACIS-S Spectrum of Mrk 3 (from Sako \etal 2000)}
\vspace{-2.7truein}
\plotfiddle{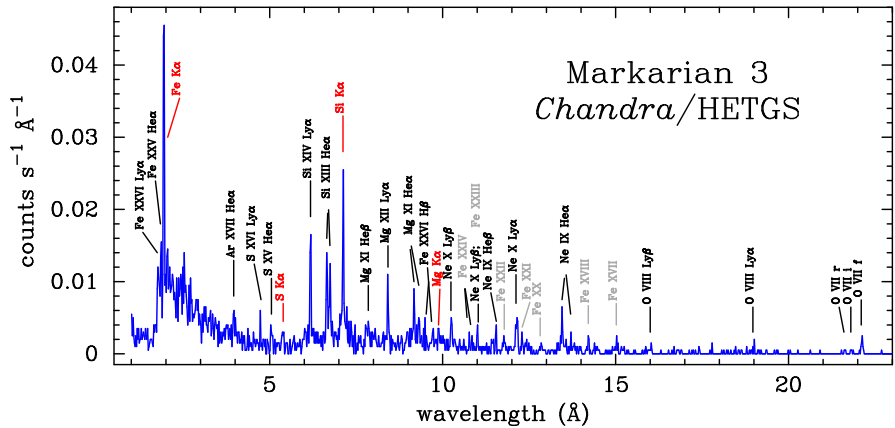}{2.5in}{-90.0}{50}{50}{-180.0}{0.0}
\vspace{2.7truein}
\end{figure}

\noindent
{\it Mrk 3}: \chandra HETG/ACIS-S observation of this Seyfert 2 galaxy
is reported by Sako \etal (2000). Figure 4 shows the beautiful
emission line spectrum with a number of lines from a variety of
elements. Resonance lines and Fe-L lines, characteristic of
photoionized plasma are strong. The OVII triplet, useful for plasma
diagnostics, is also detected. Sako \etal conclude that the emission
line plasma is clearly photoionized, with practically no contribution
from collisionally ionized gas. As such, the plasma characteristics
are consistent with a warm absorber seen in emission. Another
noteworthy result from this observation is the detection of extended
soft X-ray emission in the direction of ionization cones. This soft
emission is consistent with that due to recombination from a
photoionized plasma. \\

\begin{figure}
\caption{Extended soft X-ray emission along the ionization cone in
NGC~4151 (from Ogle \etal 2000). \chandra contours are overlayed on an
HST [OIII] 5007 image.}
\vspace{3.2truein}
\plotfiddle{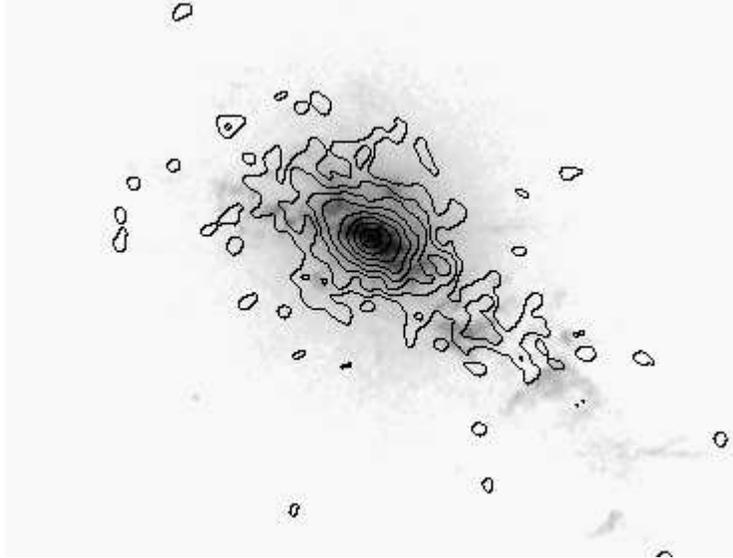}{2.5in}{0.0}{75}{75}{-220.0}{0.0}
\vspace{-2.8truein}
\end{figure}

\noindent
{\it NGC 4151}: The HETG/ACIS-S observation of this Seyfert 1.5
galaxy is reported by Ogle \etal (2000). Here again, extended soft
X-ray emission along the ionization cone is observed (Figure 5)
and numerous emission lines are detected. However, the characteristics
of the emission line plasma seem to be different from that in Mrk
3. The relative strengths of resonance and forbidden lines in the OVII
and NeIX triplet are different. This fact, together with other
features led the authors to conclude that the emission line plasma is
made of at least two distinct components: one hot, collisionally
ionized and other cooler, photoionized medium. The hot plasma can
pressure confine the cooler clouds. A large fraction of the emission
line flux seems to come from the extended emission on the scales of
the narrow emission line region. This appears to be the case even for
the Fe-K$\alpha$ line! More discussion of Fe-K lines is presented in
K. Nandra's article. \\

\noindent
The extended emission, discussed above, was known previously from {\it
Einstein} HRI and ROSAT HRI observations (e.g. NGC 4151: Elvis, Briel
\& Henry 1983; Mrk 3: Morse \etal 1995). However, the \chandra images
are spectacularly better and the additional spectral information
allows determination of physical conditions in the extended
material. See also Young \etal (2001) for \chandra observation of
extended emission in NGC~1068.

\begin{figure}
\caption{Comparison of confidence contours of the derived quantities
from \chandra and ASCA data of NGC~3783 (from Kaspi \etal 2001).}
\vspace{-2.7truein}
\plotfiddle{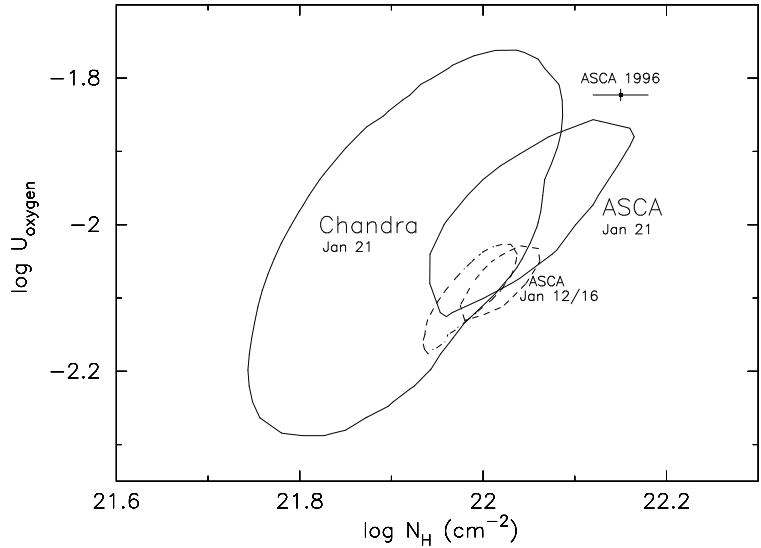}{2.5in}{-90.0}{40}{40}{-160.0}{0.0}
\vspace{3.0truein}
\end{figure}

\section{What have we learned and Where are we going?}

The data on AGNs accumulated over the first two years of \chandra are
beautiful! They show the power of the observatory both in
extraordinary imaging and high resolution spectroscopy. And they are
not just pretty pictures, or spectra. They can be tremendously useful
in gaining insights on AGN physics. However, most of the targets are
under-exposed, as is usually the case in the first year or two of
observations with a new mission. As a result, most of the absorption
line spectra have low signal to noise ratio (S/N): the measured
absorption line equivalent widths are uncertain to $\pm 40\%$. So the
keyword here is CAUTION. Figure 6 (from Kaspi \etal 2001) is
instructive: it shows that errors on derived parameters (ionization
parameter and column density in this case) from \chandra data are in
fact much larger than those from ASCA data! This clearly demonstrates
the S/N problem. So many of the spectra discussed above should be
viewed more as a proof of concept than diagnostic tools. Moreover,
most of the modeling efforts are preliminary. So once again, the
keyword is CAUTION, especially if the results are unexpected. Only
when we obtain well constrained parameters can we hope to build and
test models of AGN structure and physics. One such effort is lead by
Ian George in obtaining multiwavelength, high S/N spectra of a nearby
Seyfert galaxy with a warm absorber.  The team has observed NGC 3783
with HETG/ACIS-S for 900 ksec. This is the largest exposure for high
resolution spectroscopy of any target. Many more are sure to come in
next \chandra cycles.

However, over-emphasis on details often deprives us from having some
discovery space. We heard from the talks in other fields
(e.g. galaxies, c.f. A. Prestwich's article) that their approach was
far more exploratory. It is no surprise then that the results were
unexpected. If we do not allow for discovery space in AGN research, we
will have to depend upon serendipity for surprises and may be even for
rapid progress. Let me discuss just one example. One of the
interesting results on quasars (and there are many more; some
discussed by K. Nandra) has come from the deep field
observations. Barger \etal (2001) concluded that the quasar life times
are about half a Gyr, much larger than generally thought (incidently,
these large times are similar to what we found based on black-hole
mass to bulge mass relation of narrow line Seyfert 1 galaxies (Mathur,
Kuraszkiewicz \& Czerny 2001)). Quasar life times are intimately
related to the space density of quasars and to the question of whether
or not quasars are integral part of the life of a galaxy. Only when we
understand evolution of quasars, will we be able to give them their
rightfully important place in cosmology.

\section{Quasars as Distant Light Beams}

I will now briefly touch upon a different aspect of quasar
observations: not to study the quasars themselves but to use them as
distant light beams to look for the matter between us and the
quasar. So this is truly taking an ``X-ray'' of the Universe in
between. K. Nandra has already talked about observing the Damped
Lyman-alpha systems in the line of sight to high redshift quasars. I
will discuss the attempts to detect the warm-hot intergalactic medium
(IGM) with \chandra.

The theory of big-bang nucleosythesis together with the estimates of
deuterium abundance, imply a baryon density parameter  $\Omega_b=0.04
h_{70}^{-2}$ (Burles \& Tytler 1998). At high redshift, it appears
that most of these baryons reside in the diffuse IGM and have been
detected by Ly$\alpha$ forest absorption in background quasars
(e.g. Weinberg \etal 1997). At low redshift, however, Ly$\alpha$
forest absorption is not observed. The baryons in the stellar and
gaseous components of galaxies add up to $\Omega_b=0.004 h_{70}^{-1}$
and hot gas in clusters of galaxies makes similar contribution to
$\Omega_b$. So most of the low redshift baryons seem to be missing.

Hydrodynamic cosmological simulations predict that a large fraction of
low redshift baryons still reside in IGM, shock heated to high
temperatures ($10^5$--$10^7$ K; Cen \& Ostriker 1999), where it
produces very little hydrogen absorption.  So the most promising, and
perhaps the only way to trace the warm-hot IGM is via the ``X-ray
forest'' of high excitation metal lines (e.g. Hellston \etal
1998). For the first time in X-ray astronomy, the \chandra gratings
have enough resolution to study the narrow, intervening absorption
lines in quasars, a major field of research that has previously been
the sole preserve of optical/UV astronomy.

Fang \etal (2001) made attempts to detect the warm-hot IGM in
the line of sight to two distant quasars with HETG. These observations
did not lead to detections of intervening absorption lines. We
recently observed a nearby bright quasar H1821+643 for 500 ksec with
LETG/ACIS-S. With these observations, we expect to make the first
detection of warm-hot IGM and even the non-detection would lead to
meaningful constraints on its temperature range. So stay tuned......\\

It is my pleasure to thank the entire \chandra team for a wonderful
mission, to the organizing committee for inviting me, and to the
authors of all the papers which I have reported here. Because of the
constraints on space \& time, I could report on only a few topics and
refer to papers by only a few authors. My apologies to everybody else.


\begin{references}
\reference Bahcall, J.N. \etal 1993, ApJS, 87, 1
\reference Begelman, M., Blandford, R., \& Reed, M. 1984, Rev. Mod. Phy., 55, 255
\reference Barger, A., \etal 2001, AJ, submitted.
\reference Branduardi-Raymont, G. \etal 2001, A\&A Lett., 365, 140
\reference Burles, S. \& Tytler, D. 1998, ApJ, 499, 699
\reference Cen, R. \& Ostriker, J.P. 1999, ApJ, 514, 1
\reference Collinge, M. \etal 2001, ApJ, in press 
\reference Crenshaw D.M. \& Kraemer S.B., 1999, ApJ, 321, 233
\reference Elvis, M., Briel, U. \& Henry, P. 1983 ApJ 268, 105
\reference Fang, T., Marshall, H., Bryan, G., Canizares, C. 2001, ApJ, 555, 356
\reference Fiore, F., Elvis, M., Mathur, S.,  Wilkes, B., \& McDowell, J. 1993, ApJ, 415, 129
\reference Gallagher, S. \etal 2001, ApJ, 546, 795
\reference Gallagher, S. \etal 2001, in ``Mass Outflow in AGN: New Perspectives'', ASP Conference Series, eds. D. M. Crenshaw, S. B. Kraemer, and I. M. George
\reference George, I. \etal 2000, ApJ, 531, 52
\reference Green, P.J. \etal 2001, ApJ, in press
\reference Hellsten, U., Gnedin, N.Y. \& Miralda-Escud\'{e}, J. 1998, ApJ, 509, 56
\reference Kaastra, J. \etal 2000, A\&A Lett., 354, 83
\reference Kaspi, S. \etal 2000, ApJL, 535, 17
\reference Kaspi, S. \etal 2001, ApJ, 554, 216
\reference Krolik, J, \& Kriss, G. 1995, ApJ, 447, 512 
\reference Lee, J., \etal 2001, ApJL, 554, 13
\reference Marshall, H. L. \etal 2001, ApJL, 549, 167
\reference Mathur, S., Elvis, M. \& Wilkes, B. 1995 ApJ, 452, 230
\reference Mathur, S. 1997 in ``Mass Ejection from AGN'', Ed: N. Arav, I., Shlosman, \& R. Weymann [ASP Conf. Series Vol. 128]
\reference Mathur, S. 2000, MNRAS Lett., 314, 17
\reference Mathur, S. \etal 2000, ApJL, 533, 79
\reference Mathur, S. \etal 2001, ApJL, 551, 13
\reference Mathur, S., Kuraszkiewicz, J., \& Czerny, B. 2001, NewA, 6, 321
\reference Morse, J. \etal  1995, ApJ, 439, 121
\reference Nicastro, F., Fiore, F. \& Matt, G. 1999, ApJ, 517, 108
\reference Ogle, P. \etal 2000, ApJ, 545, 81
\reference Pradhan, A. 2000, ApJL, 545, 165
\reference Reynolds, C. \etal 1995 MNRAS, 277, 901
\reference Sambruna, R. M. \etal 2001, ApJ, 549, 161
\reference Sako, M., Kahn, S.M., Paerels, F., Liedahl, D. 2000, ApJL, 543, 115
\reference Schwartz, D. A. \etal 2000, ApJL, 540, 69
\reference Shields. J. \& Hamann, F. 1997, ApJ, 481, 752
\reference Tavecchio, F., Maraschi, L., Sambruna, R., Urry, C.M. 2000, ApJ, 544, 23
\reference Weinberg, D.H., Miralda-Escud\'{e}, J., Hernquist, L., \& Katz, N., 1997, ApJ, 490, 564

\end{references}
\end{document}